\documentstyle[aps,epsfig,floats]{revtex}

\begin{document}

\draft

\title{Long-range correlation of thermal radiation}
\author{M. Patra and C. W. J. Beenakker}
\address{Instituut-Lorentz, Leiden University, P.O. Box 9506, 2300 RA
Leiden, The Netherlands}
\date{September, 1998}

\twocolumn[
\widetext
\begin{@twocolumnfalse}
\maketitle

\begin{abstract}
A general theory is presented for the spatial correlations in the intensity of
the radiation emitted by a random medium in thermal equilibrium. We find that a
non-zero correlation persists over distances large compared to the transverse
coherence length of the thermal radiation. This long-range correlation vanishes
in the limit of an ideal black body. We analyze two types of systems (a
disordered waveguide and an optical cavity with chaotic scattering) where it
should be observable.
\end{abstract}

\pacs{PACS numbers: 42.50.Ar, 42.25.Bs, 42.25.Kb, 42.50.Lc}

\narrowtext

\end{@twocolumnfalse}
]


\begin{figure}[b!]
\epsfig{file=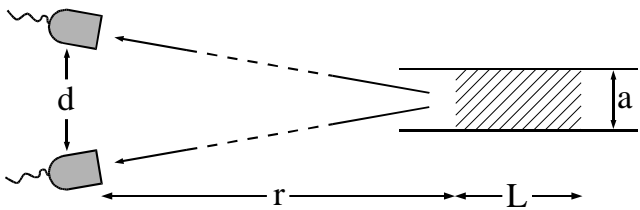,width=8.5cm}
\vspace{0.4cm}
\caption[Caption Fig. 1]{Schematic diagram of a source (length $L$,
diameter $a$)
radiating into an $N$-mode waveguide that is open at both
ends. The radiation leaving the waveguide at one end is detected by
two photodetectors at a distance $r$ from the source and separated by
a distance $d$. The photocathodes have an area below
the transverse coherence area $d_c^2\simeq r^2/N$.  We find that the
photocurrents are correlated even if the two detectors are separated
by more than $d_c$.
}
\label{aufbau}
\end{figure}

The Hanbury Brown-Twiss effect is the existence of spatial correlations
in the intensity of thermal radiation by a distant source. It was
originally proposed as an intensity-interferometric method to measure
the angular opening of a star~\cite{hanbury:56a}, far less susceptible
to atmospheric distortion than amplitude-interferometric
methods~\cite{boal:90a}. Two
photodetectors at equal distance $r$ from a source (diameter $a$) will
measure a correlated current if their separation $d$ is smaller than
the transverse coherence length $d_c\simeq \lambda r/a$ of the
radiation from the source at wavelength $\lambda$.  The correlation
function decays with increasing $d$ in an oscillatory way, with
amplitude $\propto (d_c/d)^3$~\cite{mandel:95}.

The textbook results assume that the source of the thermal radiation
is a black body, meaning that at each frequency any incident radiation
is either fully absorbed or fully reflected.
In a realistic system there will be a frequency range
where only partial absorption occurs.
The purpose of this paper is to show that in general for thermal
radiation the correlation function
does not decay completely to zero, but to a non-zero $d$-independent background
value. This long-range correlation is smaller than the short-range
correlation by a factor $(\lambda/a)^2$,
and becomes dominant for $d \gtrsim r (\lambda/a)^{1/3}$.
It contains information on deviations of the thermal radiation from
the black-body limit.

The new information contained in the long-range correlation is most easily
described when the source is embedded in a waveguide (see Fig.~\ref{aufbau}).
The waveguide has length $L$, cross-sectional area $A\simeq a^2$, and supports
$N=2\pi A/\lambda^2$ propagating modes at frequency $\omega$, counting both
polarizations. In the far-field, and near normal incidence, each mode
corresponds to a transverse coherence area $(r\lambda)^2/A\equiv d_c^2$. The
source is in thermal equilibrium at temperature $T$. The radiation emitted
through the left end of the waveguide is incident on a pair of photodetectors,
one detecting the photocurrent $I_{k}(t)$ in mode $k$, the other
detecting $I_{l}(t)$. Each photocathode has an area equal to the
coherence area or smaller.  The photocount
$n_k=\overline{n}_{k}+\delta n_{k}$ (number of photons
counted in a time $t$) and the photocurrent
$I_k=\text{d}n_k/\text{d}t=\overline{I}_{k} +\delta
I_{k}$ fluctuate around their time-averaged values
$\overline{n}_{k}$ and $\overline{I}_{k}=\overline{n}_{k}/t$.
We seek the correlation function
\begin{equation} C_{kl} =
	\int_{-\infty}^{\infty}  \overline{\delta I_k(t+\tau) \delta
	I_l(t)}\,\text{d}\tau
= \lim_{t\to\infty} \frac{1}{t}\overline{\delta
	n_k(t) \delta n_l(t)} \;.
\label{groesse}
\end{equation}
The overline
indicates an average over many measurements on the same sample.



The advantage of embedding the source in a waveguide is that we can
characterize it by a finite-dimensional
scattering matrix $S(\omega)$, consisting of
four blocks of dimension
$N \times N$,
\begin{equation}
S = \left( \begin{array}{cc} r & t \\ t' & r' \end{array} \right) \;.
        \label{smatrix}
\end{equation}
A mode $l$ incident from the left is reflected into mode $k$ with
amplitude $r_{kl}$ and transmitted with amplitude
$t'_{kl}$. Similarly, $r'_{kl}$ and
$t_{kl}$ are the reflection and transmission amplitudes for
a mode $l$ incident from the right. Reciprocity relates these
amplitudes by $r_{kl}=r_{lk}$,
$r'_{kl}=r'_{lk}$, and
$t_{kl}=t'_{lk}$.

It has been shown recently by one of the authors~\cite{beenakker:98a},
using the method of ``input-output
relations''~\cite{jeffers:93a,matloob:95a,gruner:96a}, how the
photocount distribution can be expressed in terms of the scattering
matrix. The expressions in Ref.~\onlinecite{beenakker:98a} are for a
single multi-mode photodetector.  The corresponding formulas for two
single-mode photodetectors are
\begin{eqnarray}
C_{kl} & = & \alpha_{k} \alpha_{l} \int_0^{\infty}
        |(Q Q^{\dagger})_{kl}(\omega)|^2
        \left[ f( \omega, T ) \right]^2
        \frac{\text{d}\omega}{2\pi}
       + \delta_{kl} \overline{I}_{k} \;,\nonumber\\
I_{k} & = & \alpha_{k} \int_0^{\infty}
        (Q Q^{\dagger})_{kk}(\omega)
        f(\omega,T) \frac{\text{d}\omega}{2\pi}\;,
        \label{startformel}
\end{eqnarray}
where $\alpha_{k}$ is the detector efficiency (the fraction of the
photocurrent in mode $k$ that is detected) and $f$ is the Bose-Einstein
function
\begin{equation}
	f(\omega,T)=\left[\exp(\hbar\omega/k_B T)-1\right]^{-1} \;.
\end{equation}
The $N\times N$ matrix $Q$ is related to the reflection and
transmission matrices by
\begin{equation}
        Q Q^{\dagger} = 1 - r r^\dagger -  t t^\dagger \;.
	\label{qrelatie}
\end{equation}
The integral over $\omega$ extends over a range $\Omega_c$ set by the
absorption line width, centered at $\omega_0$. Typically,
$\Omega_c\ll\omega_0$, so we can neglect the frequency dependence of
$N$ and $f$. The matrix $Q(\omega)$ for a random medium
fluctuates on a scale $\omega_c$ much smaller than $\Omega_c$. The
integration over $\omega$ then averages out the fluctuations, so that
we may replace the integrand by its ensemble average, indicated by
$\langle\ldots\rangle$,
\begin{equation}
C_{kl} = \alpha_{k}\alpha_{l}
	f^2
	\int_0^{\infty}
        \langle |(Q Q^{\dagger})_{kl}(\omega)|^2\rangle
        \frac{\text{d}\omega}{2\pi}
        + \delta_{kl} \overline{I}_{k}
        \label{cformel} \;.
\end{equation}


We evaluate the ensemble average using results
from random-matrix theory~\cite{beenakker:97a}. For a medium with
randomly placed scatterers, the ``equivalent channel
approximation''~\cite{mello:92a} has proven to be reliable. According
to this approximation, all $N$ modes are statistically equivalent.
As a consequence, for any $k\ne l$ one has
\begin{eqnarray}
\langle \text{tr} (Q Q^\dagger)^2 \rangle
	& = & N \sum_{j=1}^{N} \langle
		(Q Q^\dagger)_{kj}
		(Q Q^\dagger)_{jk} \rangle \nonumber\\
	& = & N(N-1) \langle | ( Q Q^\dagger )_{kl}
			|^2 \rangle
		+ N \langle (Q Q^{\dagger})_{kk}^2\rangle\,.
	\label{equivChannel}
\end{eqnarray}
The average of $(Q Q^{\dagger})_{kk}^2$ factorizes in the large-$N$
limit~\cite{beenakker:97a},
\begin{eqnarray}
	\langle(Q Q^{\dagger})_{kk}^2\rangle & = &
	  \langle(Q Q^{\dagger})_{kk}\rangle^2
	  \left[ 1 + {\cal O}(N^{-1}) \right] \nonumber\\
   & = & N^{-2} \langle \text{tr}\,Q Q^\dagger \rangle^2 \;.
	\label{largeN}
\end{eqnarray}
Combination of Eqs.~(\ref{equivChannel}) and (\ref{largeN}) gives us
\begin{eqnarray}
	\langle | ( Q Q^\dagger )_{kl}|^2 \rangle
	& = & N^{-2} \langle \text{tr} (Q Q^\dagger)^2 \rangle \nonumber\\
	& & {} - N^{-3} \langle \text{tr}\,Q Q^{\dagger}\rangle^2
	+{\cal O}(N^{-2}) \;.
	\label{combination}
\end{eqnarray}

The eigenvalues $\sigma_1, \sigma_2, \ldots, \sigma_N$ of the matrix
$r r^\dagger+t t^\dagger$ are the ``scattering strengths'' of the
random medium.  We denote by $\overline{\sigma^p}\equiv N^{-1} \sum_n
\sigma_n^p$ the $p$-th spectral moment of the scattering
strengths. According to Eqs.~(\ref{qrelatie}), (\ref{cformel}) and
(\ref{combination}), the cross-correlator $C_{kl}$
($k\ne l$) then takes the form of a variance,
\begin{equation}
	C_{kl} = \frac{\alpha_{k} \alpha_{l} f^2}{N}
		\int_0^\infty \left(
		\left\langle \overline{\sigma^2} \right\rangle
		- \left\langle \overline{\sigma} \right\rangle^2
		\right) \frac{\text{d}\omega}{2\pi} \;.
	\label{sigmaab}
\end{equation}
This is our basic result for the long-range correlation announced in
the introduction. The new information contained in
the cross-correlator is the variance of the scattering strengths. The
auto-correlator, in contrast, depends entirely on the first spectral
moment,
\begin{mathletters}
\label{sigmaaa}
\begin{eqnarray}
	C_{kk} & = & \alpha_{k}^2 f^2
		\int_0^\infty
		\langle 1- \overline{\sigma} \rangle^2
		\frac{\text{d}\omega}{2\pi}
		+ \overline{I}_k \;,\\
	\overline{I}_k & = & \alpha_{k} f
		\int_0^\infty
		\langle 1- \overline{\sigma} \rangle
		\frac{\text{d}\omega}{2\pi}\;,
\end{eqnarray}
\end{mathletters}
where we have used Eq.~(\ref{largeN}).

The long-range correlation $C_{kl}$ of two photodetectors
separated by more than a coherence length is an order $N$ smaller than
the short-range correlation $C_{kk}-\overline{I}_k$
of two photodetectors separated by less than a coherence length. (The
full value $C_{kk}$ is measured in a single-detector
experiment.)  The long-range correlation vanishes if all $N$
scattering strengths are the same, as they would be for an idealized
``step-function model'' of a black body ($\sigma_n=0$ for
$|\omega-\omega_0|<\Omega_c$ and $\sigma_n=1$ otherwise). A random,
partially absorbing medium, in contrast, has a broad distribution of
scattering strengths~\cite{beenakker:97a}, hence a substantial
long-range correlation of the photocurrent.

As first example, we compute the correlation for a weakly
absorbing, strongly disordered medium.
The moments of $r r^\dagger$
and $t t ^\dagger$ appearing in Eqs.~(\ref{sigmaab}) and
(\ref{sigmaaa}) have been calculated by Brouwer~\cite{brouwer:98a} as a
function of the number of modes $N$, the sample length $L$, the mean
free path $l$, and the absorption length $\xi=\sqrt{D \tau_a}$
($\tau_a$ is the absorption time and $D=c l/3$ the diffusion
constant).  It is assumed that $1/N \ll l/\xi \ll 1$, but the ratio
$L/\xi\equiv s$ is arbitrary.  The result is
\begin{mathletters}
\label{enderg}
\begin{eqnarray}
\left\langle \overline{\sigma^2} \right\rangle
		- \left\langle \overline{\sigma}
		\right\rangle^2
& = &
        \frac{2 l}{3 \xi} \left(
        \coth^3 s-\frac{3}{\sinh s}+\frac{s}{\sinh^2 s}\right. \nonumber\\
& & \left.{}+\frac{s\coth s -1}{\sinh^3 s}-\frac{s}{\sinh^4 s}\right)
        \label{enderg3} \;,\\
\langle 1- \overline{\sigma} \rangle
& = &
	\frac{4 l}{3\xi} \tanh \frac{s}{2} \;.
        \label{enderg2}
\end{eqnarray}
\end{mathletters}

\begin{figure}
\epsfig{file=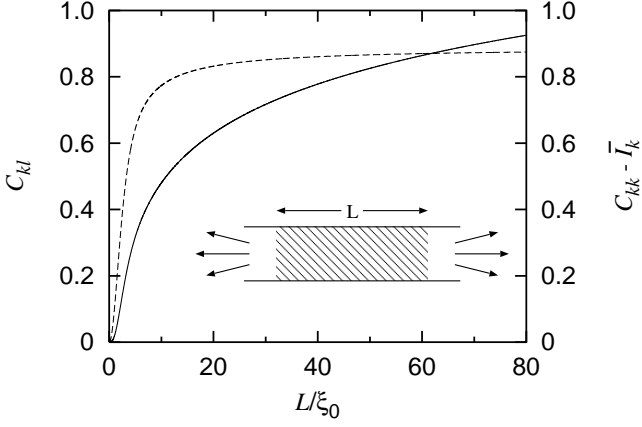,width=3.375in}
\caption{Long-range correlation
$C_{kl}$ (solid line), in units of
$\Omega_c l f^2 \alpha_k \alpha_l / N
\xi_0$,
and short-range correlation
$C_{kk}-\overline{I}_{k}$ (dashed line), in units of
$\Omega_c ( l f
\alpha_k / \xi_0 )^2$, of the radiation emitted from a disordered
waveguide (inset).
A Lorentzian frequency dependance is assumed for the dielectric function, with
width $\Omega_c$ and absorption length $\xi_0$ at the center of the absorption
line. The mean free path $l$ is assumed to be $\ll\xi_0$. The short-range
correlation saturates in the limit $L/\xi_0\to\infty$, while the long-range
correlation keeps increasing $\propto \ln L/\xi_0$.
}
\label{slabfig}
\end{figure}

To compute the correlators~(\ref{sigmaab}) and (\ref{sigmaaa}) it remains to
carry out the
integrations over $\omega$. The frequency dependence is governed by the
imaginary part of the dielectric function $\varepsilon''(\omega)$, for
which take the Lorentzian
$\varepsilon''(\omega)=\varepsilon''_0
[1+(\omega-\omega_0)^2/\Omega_c^2]^{-1}$.
Since $\tau_a=1/\omega_0 \varepsilon''$, the corresponding $\omega$-dependence
of $\xi$ and $s$ is
$\xi/\xi_0=s_0/s=[1+(\omega-\omega_0)^2/\Omega_c^2]^{1/2}$, with
$\xi_0$ and $s_0$ the values of $\xi$ and $s$ at
$\omega=\omega_0$. Results are plotted in Fig.~\ref{slabfig}. In the
limit $L/\xi_0\to0$ of a thin sample, we have
\begin{mathletters}
\begin{eqnarray}
C_{kl}&=&\frac{1}{45} \Omega_c
	\left(l f^2 \alpha_k \alpha_l / N \xi_0\right)
	\left(L/\xi_0\right)^3 \;,\\
C_{kk}&=&\frac{4}{9\pi} \Omega_c
	\left(l f \alpha_k/\xi_0\right)^2
	\left(L/\xi_0\right)^2 + \overline{I}_k \:,\\
\overline{I}_k&=&\frac{1}{3} \Omega_c
	\left(l f \alpha_k/\xi_0\right)
	\left(L/\xi_0\right) \:.
\end{eqnarray}
\end{mathletters}
In the opposite limit $L/\xi_0\to\infty$ of a thick sample, the
cross-correlator $C_{kl}$ and the mean current
$\overline{I}_{k}$ both diverge logarithmically $\propto \ln
L/\xi_0$. The ratio $C_{kl}/(\overline{I}_k
\overline{I}_l)^{1/2}$ tends to $(1/2 N) f \sqrt{\alpha_k
\alpha_l}$ in the large-$L$ limit, and
the short-range correlation
$C_{kk}-\overline{I}_{k}$ to
$\frac{8}{9}\Omega_c (l f \alpha_k/\xi_0)^2$,
remaining larger than the long-range correlation because the limit
$N\to\infty$ has to be taken before $L\to\infty$.

\begin{figure}[b!]
\epsfig{file=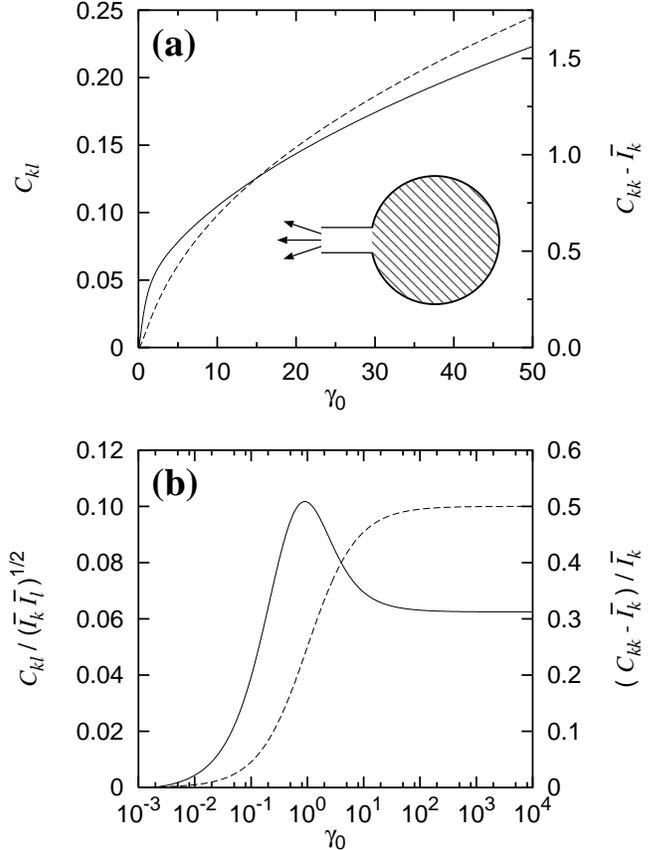,width=3.375in}
\caption[Caption for quantum dot]{Correlators of the radiation emitted from a
disordered optical cavity (inset) as a function of the absorption rate
$\gamma_0$
at the center of
the absorption line with Lorentzian profile. (The absorption rate is normalized
to the mean dwell time.)
{\bf (a)}~
Long-range correlation
$C_{kl}$ (solid line), in units of $\Omega_c f^2 \alpha_{k}
\alpha_{l} / N$,
and short-range correlation
$C_{kk}-\overline{I}_{k}$ (dashed line), in units of $\Omega_c
f^2 \alpha_{k}^2$.
{\bf (b)}~Same correlators, but now normalized by the mean photocurrent. (The
left axis is in units of
$f \sqrt{\alpha_k \alpha_l}/N$, the right axis in units
of $f \alpha_k$.) The long-range correlation persists in the limit
$\gamma_0\to\infty$ because of partial absorption in the tails of the
absorption
line.
}
\label{qdotfig}
\end{figure}

Our second example is an optical cavity
filled with an absorbing random medium (see Fig.~\ref{qdotfig}a,
inset). The radiation leaves
the cavity through a waveguide supporting $N$ modes. The general
formula~(\ref{startformel}) applies with $Q Q^\dagger=1-r r^\dagger$
(since there is no transmission).
The scattering strengths $\sigma_1, \sigma_2, \ldots, \sigma_N$ in
this case are eigenvalues of $r r^\dagger$. Their distribution is
known in the large-$N$ limit~\cite{beenakker:98b} as a function of the
dimensionless absorption rate $\gamma=2\pi/N\tau_a \Delta\omega$, with
$\Delta\omega$ the spacing of the cavity modes near frequency
$\omega_0$. (The quantity $\gamma$ is the ratio of the mean dwell time in the
cavity without absorption and the absorption time.)
The moments
$\langle\overline{\sigma}\rangle$ and
$\langle\overline{\sigma^2}\rangle$ can then be computed by numerical
integration.
Results are shown in Fig.~\ref{qdotfig}, again for a Lorentzian
frequency dependence of $\varepsilon''(\omega)$. Unlike in the first
example, we are now not restricted to weak absorption but can let the
absorption rate $\gamma_0$ at the
central frequency $\omega_0$ become arbitrarily large.
For weak absorption, $\gamma_0\ll 1$, we have
\begin{mathletters}
\label{qdotklein}
\begin{eqnarray}
C_{kl}&=&\case{1}{4} \Omega_c
        \left( f^2 \alpha_k \alpha_l / N \right)
        \gamma_0^2 \;,\\
C_{kk}&=&\case{1}{4} \Omega_c
        ( f \alpha_k \gamma_0)^2 + \overline{I}_k,\quad
\overline{I}_k = \case{1}{2} \Omega_c
        f \alpha_k
        \gamma_0 \;.
\end{eqnarray}
\end{mathletters}
For strong absorption, $\gamma_0\gg 1$, all three quantities
$C_{kl}$, $C_{kk}$ and $\overline{I}_k$ diverge
$\propto \sqrt{\gamma_0}$ (see Fig.~\ref{qdotfig}a). The ratio
$C_{kl}/(\overline{I}_k\overline{I}_l)^{1/2}$ tends to
$0.062 f (\alpha_k \alpha_l)^{1/2} /N$, and the ratio
$(C_{kk}-\overline{I}_k)/\overline{I}_k$ to $\frac{1}{2}
f \alpha_k$ (see Fig.~\ref{qdotfig}b).
The long-range correlation
does not vanish as $\gamma_0\to\infty$, because there remains a
tail of frequencies with moderate absorption and thus a wide
distribution of scattering strengths, even if the system behaves like
an ideal black body for frequencies near $\omega_0$.


In summary, we have shown that the thermal radiation emitted by
random media contains long-range spatial correlations in the intensity. The
long-range correlation has information on the spectral variation of the
scattering strengths that is
not
accessible from the luminosity. We have analyzed two types of systems in
detail,
providing specific predictions that we hope will motivate an experimental
search
for the long-range correlation.

This work was supported by the Dutch Science Foundation NWO/FOM.

\end{document}